\documentstyle[12pt,a4]{article}

\begin{document}

\begin{center}
{\large \bf Excitons in type-II quantum dots: Finite offsets}
\end{center}

\vspace*{3cm}

\begin{center}
U.\ E.\ H.\ Laheld, F.\ B.\ Pedersen, and P.\ C.\ Hemmer\\

Institutt for fysikk, Norges tekniske h\o gskole,\\
N-7034 Trondheim, Norway

\end{center}

\vspace{1cm}

\noindent
{\em Abstract:}\\
Quantum size effects for an exciton attached to a spherical quantum
dot are calculated by a variational approach. The band line-ups are
assumed to be type-II with
finite offsets $V_e$ and $V_h$. The dependence of the exciton binding
energy
upon the dot radius $R$ and the offsets is studied for different sets
of electron and hole effective masses.

\newpage

\noindent{\bf I. INTRODUCTION}\\

With the development of new techniques for fabrication of semiconductor
heterostructures, quantum-size effects in these low-dimensional
structures
(quantum wells\cite{wells,wells1,wells2,wells3,wells4,wells5,wells6},
 wires\cite{wires} and
dots\cite{dots,dots1}) have been studied extensively\cite{Yoffe}. The
quantum state of an Wannier exciton is one main subject in this
respect,
since, by the spatial confinement in these microstructures, it is
qualitatively different from the exciton state in bulk materials.

Several theoretical studies, with various degrees of sophistication,
exist for excitons in type-I heterostructures for quantum wells,
quantum wires and quantum dots. Excitons in type-II heterostructures
have also been studied\cite{dots,dots1,Duggan,Branis}, but to a much
lesser extent. A fundamental feature of type-II heterojunctions is the
spatial separation of electron and hole, which leads to longer
radiative lifetimes, lower exciton binding energy, and unusual
dynamic and recombination properties of charge carriers as compared
with type I. For an overview, including potential applications, we
refer to a recent article\cite{Mikhail}, from which we quote:
``Though some III-V, IV-VI and II-VI semiconductor materials can
form type II junctions (AlInAs/InP, InAsSb/InSb, InAs/GaSb,
GaInAsSb/GaSb, InGaAs/GaAsSb, Si/Ge, ZnTe/ZnSe, etc) the
intriguing properties of these remarkable structures are still poorly
understood.''

 The case of
excitons in type-II quantum dots has only recently been considered in
two model calculations.
Rorison\cite{R} has used a simple separable wave function in a
variational calculation, with parameters appropriate for GaAs/AlAs
and InAs/GaSb dots.
We\cite{LPH} have, on the other hand, used a more sophisticated
variational wave
function, and also presented analytical   considerations of limiting
cases, with the aim to obtain insight in how  the
exciton binding energy and the electron-hole correlations
depend upon the effective masses and the dot radius.
 In Ref.\ \cite{LPH}
the offsets were assumed to be infinite, corresponding to a complete
spatial separation of  the electron and hole. It was shown that two
different regimes exist: For dot radii $R$ much smaller than the
effective
bulk Bohr radius the electron and hole are essentially  uncorrelated,
while for $R$ much larger than the Bohr radius the electron and hole
are strongly spatially correlated, residing near the
dot boundary just opposite each other.

 This infinite barrier model
is artificial for small dot radii $R$, since in realistic situations
the confined particle then tends to leak out in the barrier material.
Moreover, the importance of the incomplete confinement in optical
experiments is very clear, since the oscillator strength for
excitonic transitions is proportional to the square of the
electron-hole
wave-function overlap. Thus it will be worthwhile to make clear the
dependence of the exciton binding energy, and of the wave function
overlap, on the magnitudes of the offsets.
For estimates of band offsets and effective masses for several III-V
heterostructures see Refs.\ \cite{Mikhail,Krijn,Tiwari}.

In this work, we report the result of a variational calculation of
the binding energy for the electron-hole system with {\em finite}
offsets.  Since it is always useful to have results for limiting
cases to compare with,
we compute first   binding energies for the case in which one particle
is confined, and the other completely free, i.\ e.\
$V_h = 0$.  Since zero offset for one of the particles
is a situation intermediate between a type-I and a type-II
heterostructure,
we denote this special case as being of type I$\frac{1}{2}$.
A limiting type-I$\frac{1}{2}$ situation occurs when the confined
particle is {\em completely} confined within the dot ($V_e = \infty$).
In Sec.\ III the dependence of the exciton binding energy upon the dot
radius, the offsets and the effective-mass values  is studied, while
Sec.\ IV contains results for the electron-hole overlap in the
wave function, an important factor for the magnitude of oscillator
strengths.
 We summarize our findings in Sec.\ V.\\

 \noindent
{\bf II. MODEL AND METHOD}\\

In the effective-mass treatment the Hamiltonian for the
electron-hole pair which forms the exciton is given by
\begin{equation}
H =  \frac{p_e^2}{2m_e} + \frac{p_h^2}{2m_h} -
\frac{e^2}{4\pi \epsilon}
\frac{1}{|{\bf r}_e-{\bf r}_h|} + V_e({\bf r}_e) + V_h({\bf r}_h),
\label{H}
\end{equation}
where

\begin{equation}
V_e({\bf r}) = \left\{ \begin{array}{cc}
                          0  & \mbox{for }\; r\leq R\\
                          V_e& \mbox{for }\; r>R
                          \end{array}
               \right.
\label{Ve}
\end{equation}
and

\begin{equation}
V_h({\bf r}) = \left\{ \begin{array}{cc}
                          V_h  & \mbox{for }\; r\leq R\\
                          0 & \mbox{for }\; r>R .
                          \end{array}
               \right.
\label{Vh}
\end{equation}

Type-II situations correspond to $V_e$ and $V_h$ having the same sign.
For definiteness we assume both positive so that the electron is
 the confined particle.
The alternative configuration
presents of course the same computational problem.

For simplicity we have assumed the same electron (and hole) effective
mass in the dot and the barrier material, and
 that the dielectric constants of the two media can be accounted
  for by a single average value $\epsilon$.   With
degenerate valence band, we let  $m_h$ be either the heavy-hole or
the light-hole effective mass, thus neglecting the complications due
to the off-diagonal terms in the Kohn-Luttinger Hamiltonian
\cite{gaute}.

The main task is now to determine the ground-state energy $E_0$ of the
Hamiltonian (~\ref{H}). The binding energy $E_b$ of the exciton is then
the energy required to remove the hole,
given by the difference between $E_0$ and the confinement energy
$E_c $,  the ground-state energy of the electron in the spherical dot:
\begin{equation}
E_b = E_c - E_0,
\label{bind}
\end{equation}.

As units of energy and length we use an effective Rydberg energy
\begin{equation}
E_{Ry}^h = \frac{m_h}{2\hbar^2} \left(\frac{e^2}{4\pi\epsilon}\right)^2
\label{Rydberg}
\end{equation}
and an effective Bohr radius
\begin{equation}
a_h = \frac{4\pi \epsilon \hbar^2}{m_he^2},
\end{equation}
both in terms of the hole mass. In terms of these quantities we denote
\begin{equation}
\hat{E} = \frac{E}{E_{Ry}^h}
\end{equation}
as our dimensionless energy, and
\begin{equation}
\hat{R} = \frac{R}{a_h}
\end{equation}
as the dimensionless radius.

The ground-state energy of the Hamiltonian (~\ref{H}) is determined
variationally, using the following nonorthogonal basis set of functions:
\begin{equation}
\psi({\bf r}_e,{\bf r}_h) =\sum_{k=1}^{N_e} \sum_{l=1}^{N_h} C_{kl}\;
e^{-k\eta r_e^2} \;e^{-l\gamma r_h}\; e^{-\beta|{\bf r}_e-{\bf r}_h|}.
\label{var}
\end{equation}
The variational parameters are $\eta$, $\gamma$, $\beta$, and
the expansion coefficients $C_{kl}$, altogether $3+N_eN_h$ parameters.
 Optimalization with respect to
the expansion coefficients $C_{kl}$ is a generalized eigenvalue problem,
consisting in diagonalization of a matrix of size $N_e N_h \times
N_e N_h$.  All matrix elements can be evaluated analytically in terms
of error functions.
The only numerical problems that need special attention arise
 because the analytic expressions contain
terms that individually diverge when $\gamma_k+\gamma_l-2\beta$
is close to zero, or because  exponential terms  may become
inadmissably large for small $\eta$s.
Finally, the minimalization
with respect to $\eta$, $\gamma$ and $\beta$ is demanding due
to the existence  of many local minima.

Good accuracy is usually
achieved with a moderate size of the basis set. This is
illustrated by Table 1, which
shows, by means of a special example, how the result for the
ground-state energy of the exciton depends upon the size of the
basis set.
Apparently a basis set of 9 functions ($N_e = N_h = 3$)
suffices to give the energy within 1\% accuracy.  The exceptional
cases occur for very large offsets, because one cannot achieve an
almost vanishing wave function in the nearly inaccessible region
with a small number of
basis functions, and for large dot radius $R$ when the exciton
is located near
the boundary. In these special cases we make use of  alternative
functional forms, as discussed below.

One may of course also check the quality of the Gaussian basis set for
the confined particle by comparing the one-electron energies computed
variationally with the known exact ground state. With $N_e =5$, the
accuracy is in general on the 1\% level, or better.
Although the confinement energy $E_c$ in equation (~\ref{bind}) may
be obtained exactly,  we compute it variationally for better
procedural consistency.     \\

\noindent
{\bf III. EXCITON BINDING ENERGY}\\

We first discuss the case of one particle confined to the dot material,
 with the other free to move. We denote this situation, intermediate
between a type-I and a type-II heterostructure, as type I$\frac{1}{2}$.
Two examples of heterostructures with very small valence-band offsets
are InAs/AlSb \cite{Dow} and ZnSe/Zn$_{0.79}$Mn$_{0.21}$Se \cite{Citrin}.

  A negligible hole offset, and a {\em completely}
confined electron constitutes the extreme case:\\

\noindent
{\bf A. $V_e = \infty$, $V_h = 0$}\\

For this extreme case we use, for the reason mentioned above,
 an alternative variational function
\begin{equation}
\psi({\bf r}_e,{\bf r}_h) =
\frac{\sin(\pi r_e/R)}{r_e}\; e^{-\gamma r_h}\;
e^{-\beta |{\bf r}_e -{\bf r}_h|}
\label{var2}
\end{equation}
for $ r_e \leq R$, zero otherwise.
The numerical results for the exciton binding energy $E_b$ are shown
 in figure ~\ref{fig1}.    Fig.\ ~\ref{fig2} exhibits the
 size effect of the three different contributions to the
total energy, viz.\ the kinetic energies of the electron and hole and
the Coulomb interaction energy.  We see that for $\hat{R}$ smaller
than about
3 the electron kinetic energy is essentially equal to the ground-state
energy in
the  dot, and dominates the other energy contributions. For large
$\hat{R}$
the three energy contributions approach the ratio 1:1:-4,
characteristic of the two-particle problem in bulk.

It is  possible to understand the large-$R$ and small-$R$ behavior of
the binding energy in figure ~\ref{fig1}. For large radii the binding
energy must approach
the bulk binding energy with a {\em reduced} mass:

\begin{equation}
\hat{E}_b(R\rightarrow \infty) = \frac{m_e}{m_e+m_h}.
\end{equation}

For small enough dot radius, or small electron to hole mass ratio
($\hat{R}^2m_e/m_h << 1$), the electron
will reside in its ground state $\psi_0(r_e)=
 \sin(\pi r_e/R)/(r_e\sqrt{2\pi R})$. Then the hole will see a
 charge distribution $\rho(r)= -e\psi_0^2(r)$, corresponding to a
 potential
\begin{eqnarray}
V(r)& =& \int \frac{e\rho(r_e)}{4\pi \epsilon
|{\bf r}-{\bf r}_e|} d^3r_e   \nonumber \\
& = &\left\{ \begin{array}{l}
-\frac{e^2}{4\pi \epsilon \; r}
 \hspace*{1cm}  \mbox{for $r>R$} \\
-\frac{e^2}{4\pi \epsilon  R}
 [1-\frac{R}{2\pi r}\sin \frac{2\pi r}{R} +
2\int\limits^1_{r/R} \sin^2(\pi x) \frac{dx}{x} ]  \\
\hspace*{1cm} \mbox{for $r<R$}
\end{array}
\right.
\label{V}
\end{eqnarray}
The radial Schr\"{o}dinger equation with this potential gives a
 ground state
that is in good numerical agreement with the $m_e/m_h = 0$ curve of
figure ~\ref{fig1}. For a finite effective-mass ratio this
treatment can legitimately be used for small $\hat{R}$ only,
in which case the potential (~\ref{V}) is  merely a small perturbation
of the Coulomb potential. Using the Coulomb potential as the
unperturbed potential and the difference $V(r)+e^2/4\pi\epsilon r$
as the perturbing  potential first-order perturbation theory yields
\begin{equation}
\hat{E}_b = 1 - {\textstyle \left(\frac{4}{9}-\frac{2}{3\pi^2}\right)}
 \hat{R}^2 +  {\cal O}(\hat{R}^3).
 \end{equation}
For equal effective masses, it can be checked numerically that
this is a reasonable approximation for
$\hat{R}<0.5$. This range of dot radii is so small that the
parabolic top is not visible in figure ~\ref{fig1}.

The spatial correlation function
\begin{equation}
C=\frac{\langle({\bf r}_e-\langle{\bf r}_e\rangle)
    ({\bf r}_h-\langle{\bf r}_h\rangle) \rangle}{\sqrt{\langle (
    {\bf r}_e-\langle {\bf r}_e \rangle )^2\rangle  \langle (
    {\bf r}_e-\langle {\bf r}_e \rangle )^2\rangle }}  =
\frac{\langle{\bf r}_e {\bf r}_h \rangle}{\sqrt{\langle r_e^2 \rangle
      \langle r_h^2 \rangle }}
\label{corr}
\end{equation}
gives a quantitative measure of correlations. Figure ~\ref{fig3}
shows the results. For small $m_e$, as well as for a small dot radius,
the particles are weakly correlated, since the electron is
essentially in its ground state. And, as could be expected, the
 correlations are larger in the present case than for an infinite
hole offset.

For larger electron mass the correlations increase. The limiting case
of a very large mass ratio ($m_e >>m_h$) can easily be treated in the
Born-Oppenheimer approximation, which yields  the limiting
behavior\cite{detail}
\begin{equation}
C(m_e/m_h \rightarrow \infty) \rightarrow
\frac{\hat{R}}{\sqrt{\hat{R}^2+9/(1-{\textstyle\frac{3}{2}}\pi^{-2})}}.
\label{C}
\end{equation}
The correlation function for the largest mass ratio in figure

\noindent
{\bf B. $V_e$ finite, $V_h = 0$}\\

For finite electron offset, still in a type-I$\frac{1}{2}$ situation,
we obtain more accurate results by means of
the variational function (~\ref{var}). The results for equal effective
masses are given in figure
have a minimum size in order to be able to bind the electron after
the break-up of the  exciton. The presence of the hole, however, which
in
type-I$\frac{1}{2}$ situations merely is a satellite to the electron,
makes it possible to have the exciton attached to smaller dots, dots
for
which the offset potential is insufficient to bind the single
electron\cite{LH}.

The figure shows that the binding is weaker when the dot is large, and
when the offset is low. All graphs show that for a given offset there
exists a radius $R_m$ for which the binding energy is maximal, and
$R_m$ increases with decreasing $\hat{V}_e$.
 This trend can be understood
by envisaging the confined particle to create a charge distribution,
with which
the freely moving particle then interacts. The ground state in a
smeared
charge distribution deviates from the Coulomb value increasingly more
 the more extended the charge distribution is since the overlap
 decreases. Our interpretation is that the maximum binding energy of
the two-particle system corresponds to the minimum extension of the
wave function of the confined particle. Since the
 one-particle ground state in
a spherical dot is a well-known textbook example, one can calculate
 its width
$\langle r^2 \rangle^{\frac{1}{2}}$ exactly. The width is very
large both when the dot is very large, and when the dot is so small
that the potential barely binds the electron. Thus a {\em minimum}
width  of the one-particle ground state
must exist for a definite dot radius $r_m$, and a straightforward
calculation\cite{detail} yields the exact behavior
\begin{equation}
\hat{r}_m = c  \sqrt{\frac{m_h}{m_e}} \; \sqrt{\frac{1}{\hat{V}_e}},
\label{min}
\end{equation}
with a numerical constant $c = 2.67$. For the offsets $\hat{V}_e = 1,
 5$
and $25$ in figure ~\ref{fig4} the binding energy
 is maximal at dot sizes $\hat{R}_m = 2.35, 1.11$
and 0.53, respectively. This corresponds to the values 2.35, 2.48 and
2.65, respectively,    for
$\hat{R}_m \sqrt{\hat{V}_e}$, showing that the interpretation
of the maxima makes sense, for large offsets even quantitatively.

The binding energy also  depends upon the effective masses. As shown in
figure ~\ref{fig5}, where $\hat{V}_e=25$, the binding energy decreases
when the electron
effective mass becomes smaller.
 The position of the maximum of $\hat{E}_b$
is seen to increase with decreasing effective mass ratio $m_e/m_h$
in close accordance with the square-root dependence of
Eq.\ (~\ref{min}).
Numerically, the maxima in figure ~\ref{fig5} occur at dimensionless
radii 1.63, 1.11, 0.71, and 0.53 for the mass ratios $m_e/m_h$ =
0.1, 0.2, 0.5,
and 1, respectively, while Eq.\ (~\ref{min}) yields the values
$\hat{r}_m $ = 1.69, 1.19, 0.76, and 0.53 for these mass ratios.

Let us consider a specific case.
As mentioned above InAs/AlSb is close to a type I$\frac{1}{2}$
heterostructure. We quote from Ref.\ \cite{Dow}: ``Since the valence
band offset of 0.11 eV between InAs and AlSb is so small, we may
approximate it as zero, ...''. The electron, on the other hand,
is well confined in the InAs material, since $V_e = 1.37$eV
\cite{Tiwari}. Using\cite{Landolt} an average dielectric constant
$\epsilon \simeq 13.6 \epsilon_0$, an average light-hole effective
mass $m_{h,l} \simeq 0.07 m_0$,
 we find values of the dimensionless electron
offset $\hat{V}_e$ to be more than 250, i.\ e.\ effectively infinite.
Taking the average heavy-hole effective mass to be
$m_{h,h} \simeq 0.5 m_0$, we find
$\hat{V}_e \approx 35$, also very large.
The relevant figure is thus Fig.\ 1. With the electron effective mass
in InAs, $m_e = 0.022 m_0$,
  the mass ratios $m_e/m_h$ are about 0.3 and 0.04,
respectively.  One thus have to interpolate between the mass ratios
0 and 0.5 graphs in Fig.\ 1.  (For the heavy-hole exciton this can only
be used for dimensionless dot radii larger than about 2.3, according to
the expression (~\ref{min}).  The lower curve in Fig.\ 5,
although corresponding to an offset
25 rather than 35,  and a larger mass ratio $m_e/m_h$, gives also a rough
idea of how the binding energy for the heavy-hole exciton
 depends on the dot radius.)      \\

 \noindent
{\bf C. $V_e$ finite, $V_h$ finite}\\

We finally investigate the effect of a finite hole offset. The binding
energy as function of the quantum dot radius $R$ is shown in figure
Comparison with figure ~\ref{fig4} shows that the binding energy is
smaller
when the hole offset is nonzero, as could be expected. We  see,
moreover,  that
offsets larger than about 15 yield essentially the same binding
energy as infinite offsets.

 The figure corresponds to $m_e=m_h$. The
binding energy for equal electron and hole offsets is, however, very
insensitive to the mass ratio $m_e/m_h$: Decreasing the mass ratio from
1 to 0.5 lowers $\hat{E}_b$ at most by a few percent. This insensitivity
was also found for infinite offsets \cite{LPH}.  However, for a large
disparity between the electron and hole offsets, the binding energy is
more sensitive to the mass parameters, as witnessed by
figure ~\ref{fig5}.

The figure
also shows the interesting feature that with increasing values of
 $\hat{V}_e$, $\hat{E}_b$ now {\em decreases},
 while it {\em increased} in the absense of the hole offset
(Fig.\ ~\ref{fig4}). How can this be explained?
An increase of any of the two offsets is a positive
 definite perturbation on the Hamiltonian (~\ref{H}), and leads
  therefore to
an increase in its ground-state energy $E_0$, i.e.\ gives a negative
contribution to the binding energy $E_b$. Hence $E_b$ would be lowered if
the hole offset were increased for fixed $V_e$. An increase in the
{\em electron} offset, however, has the additional effect of increasing
the electron confinement energy $E_c$  in equation (~\ref{bind}), and
the competition between the two effects determines the change in
the binding energy $E_b=E_c-E_0$. The variation of the confinement
energy $E_c$
is the dominating contribution when $V_h=0$ (Fig.\ ~\ref{fig4}),
but cannot outweigh the combined influence on $E_0$ from
increasing {\em both} the hole and electron offsets by equal amounts
(Fig.\ ~\ref{fig6}).

For type-II heterostructures the offsets are equal if the
bandgaps in the two materials are equal. For GaAs and InP the bandgaps
are (at 300K) 1.42 eV and 1.34 eV, respectively\cite{Landolt}.
 The dielectric constant in both
materials is close to $\epsilon = 12.7 \epsilon_0$.
The band offsets are small, estimated\cite{Nolte} to be of order 0.2 eV
and 0.3 eV. Let us assume that this may be approximated by a
common offset value of 0.25 eV, and let us assume further that the
complications due to to strain can be neglected.
 With an
average electron effective mass $m_e \simeq 0.07 m_0$, and average
light-hole effective mass $m_{h,l} \simeq 0.10 m_0$, and the average
heavy-hole effective mass $m_{h,h} \simeq 0.54 m_0$, we obtain the
very rough estimates 30 (light hole) and 5 (heavy hole) for the
dimensionless offset values. Thus the exciton binding
energy versus radius graph for offset value 5 in Fig.\ 6 is
relevant for the heavy-hole exciton,
and the graph corresponding to the light-hole exciton is squeezed
between the two proximate graphs for offset values 15 and infinity.
As noted above the results are insensitive to the value of
the mass ratio.                      \\

\noindent
{\bf IV. Oscillator strengths}\\

The hole-electron overlap is a decisive factor in determining the
properties of the exciton in optical experiments. Since infinite
offsets give zero overlap,
the magnitudes  of the offsets must clearly be extremely important
in type-II situations.

The oscillator strength of an optical transition is proportional to
the factor\cite{HN}
\begin{equation}
f_0 =\left|\int \psi({\bf r},{\bf r}) \;d^3r\right|^2.
\label{overlap}
\end{equation}
Figure ~\ref{fig7} shows how the overlap factor $f_0$ increases
dramatically with decreasing offsets.  This is expected, of course,
since the particles are able to penetrate the dot boundary
when the offsets are lowered.
In figure ~\ref{fig8} we illustrate a typical situation, for
 dimensionless offsets equal to unity, and dimensionless
dot radius $\hat{R} = 8$.  The marginal
electron and hole distributions, obtained from the two-particle
wave function, are clearly centered on each side of, and away from,
the dot boundary.  The tails of these marginal distributions
are small {\em at} the boundary, while the overlap, which corresponds
to ${\bf r}_e= {\bf r}_h$, is maximal at, or very close to, the
dot boundary.

How the overlap factor depends upon the size of the quantum dot
is more interesting.
The electron wave function is squeezed out
of the dot volume when the radius becomes small enough, just as
it is for small  offsets.
Then the electron and hole are able to correlate, yielding a large
overlap function.
The decrease of the overlap function with increasing $R$, shown in
 figure ~\ref{fig9}, is in accordance with this. However, $f_0$ must
finally increase with $R$ as
\begin{equation}
f_0 \simeq c\cdot 4\pi \hat{R}^2,   \mbox{\hspace*{2cm} $\hat{R}
\rightarrow \infty$}
\label{asym}
\end{equation}
where the constant $c$ depends on the offsets and on
the effective-mass
ratio.  The rationale behind (~\ref{asym}) is that for very large
radii the exciton will, on the scale of the dot radius,
 sit very close to the boundary, so that the determination of
 the wave function is
essentially determined through the solution of a plane-wall problem.
The plane-wall problem determines the
constant $c$, and the additional factor of $4\pi \hat{R}^2$ in
(~\ref{asym})
comes from normalization and integration over the dot surface.
For the exciton near a plane wall at $z=0$ we use the simple
 variational function
$\psi({\bf r}_e,{\bf r}_h) = (C_e-z_e)(C_h+z_h)
e^{-\beta|{\bf r}_e-{\bf r}_h|}$ for $z_e<C_e, \;\;z_h>-C_h$,
and zero otherwise.
For the parameters of Fig.\ ~\ref{fig9} we find for the
variational parameters $C_e=C_h= 0.85   $ and $\beta = 0.15  $,
which implies
$c = 1.62 \cdot 10^{-7} $. The rise of $f_0$ according
to (~\ref{asym}) is outside the range of radii shown in the figure.\\

\noindent
{\bf V. Summary}\\

The present work is a model calculation in which we have,
 in the effective-mass treatment, performed a
variational calculation of excitons in spherical quantum dots.
We have in particular studied how the exciton binding energy $E_b$
depends
on the dot radius $R$, the effective masses $m_e$ and $m_h$, and the
offsets. The nature of the binding energy maximum, as function of
the dot radius, is clarified.

\noindent
{\bf Acknowledgements}\\

We thank G.\ T.\ Einevoll for useful remarks. Two of us (U.\ E.\ H.\
Laheld and F.\ B.\ Pedersen) are grateful to Norges Forskningsr\aa d
for support.

\newpage

\begin{figure}
\caption[]{The dimensionless exciton binding energy $\hat{E}_b$
for the offsets $V_e =\infty$,
$V_h=0$, as function of the dimensionless dot radius $\hat{R} = R/a_h$.
 The
values of the mass ratio $m_e/m_h$ are shown on the right-hand side.}
\label{fig1}
\end{figure}

\begin{figure}
\caption[]{The electron kinetic energy (e), the hole kinetic energy (h),
and the absolute value of the Coulomb energy (c), as functions of the
dimensionless quantum dot radius $\hat{R}$. These three contributions
to the total energy are measured in the effective Rydberg
(5).
Here $m_e = m_h$, $V_e = \infty$, and $V_h = 0$. The dashed line
corresponds
to the ground state of the single electron.}
\label{fig2}
\end{figure}

\begin{figure}
\caption[]{The spatial correlation function (30) as a function of the
dimensionless radius $\hat{R}=R/a_h$ of the quantum dot, for the case
$V_e =\infty$. The fully drawn curves correspond to $V_h=0$, the dashed
curves to $V_h=\infty$ [16]. The two upper graphs
correspond to
$m_e=100m_h$, the two lower ones to $m_e=m_h$.}
\label{fig3}
\end{figure}

\begin{figure}
\caption[]{The dimensionless exciton binding energy $\hat{E}_b$   for
$V_h=0$ and several values of $V_e$, as function of the
dimensionless quantum dot
radius $\hat{R}$.  The values of $\hat{V}_e$ are given on the
graphs. Here $m_e = m_h$. In
the variational calculation $N_e =5$ and $N_h =2$ have been used.
 The dashed line corresponds to $V_e=\infty$,
calculated less accurately with
the simpler function (10).}
\label{fig4}
\end{figure}

\begin{figure}
\caption[]{The dimensionless exciton binding energy $\hat{E}_b$ for
$V_h=0$ and $\hat{V}_e = 25$, as function of the dimensionless
quantum dot radius $\hat{R}$, for several values of the effective mass
ratio $m_e/m_h$ (shown on the right-hand side).  In the variational
calculation $N_e=5$ and $N_h=2$ have been used.}
\label{fig5}
\end{figure}

\begin{figure}
\caption[]{The dimensionless exciton binding energy $\hat{E}_b$ for
equal effective masses and equal offsets, as function of the
dimensionless quantum dot radius $\hat{R}$. The offset values are
shown on the graphs. In the variational calculation
$N_e=3$ and $N_h=5$ have been used. Results corresponding to
$V_e = V_h=\infty$ have been taken from Ref.\ [16]
 and are shown as a dashed graph.}
\label{fig6}
\end{figure}

\begin{figure}
\caption[]{The overlap factor (17) as function of
the offset.
Here $m_e=m_h$, $\hat{V}_e = \hat{V}_h =\hat{V}$, and $\hat{R}=3$.
 In the variational calculation
$N_e=3$ and $N_h=5$ have been used.}
\label{fig7}
\end{figure}

\begin{figure}
\caption[]{The radial probability densities
$4\pi r_i^2 \psi(r_i)^2$, $i=e,h$
for the electron position (e) and
the hole position (h), as functions of the dimensionless
distance from the dot
center.  Here $\hat{R} = 8$ (indicated by an arrow),
 $m_e=m_h$ and $\hat{V}_e=\hat{V}_h =1$.
The dashed line shows $10^4 4\pi r^2 \psi({\bf r}, {\bf r})^2$.
(The small wiggle near
$\hat{r}=3$ is presumably due to numerical inaccuracies.)
In the variational calculation
$N_e=3$ and $N_h=5$ have been used.}
\label{fig8}
\end{figure}

\begin{figure}
\caption[]{The overlap factor (17) as function of the
dimensionless dot radius $\hat{R}$.  Here $m_e=m_h$ and
 $\hat{V}_e=\hat{V}_h= 1$. In the variational calculation
$N_e=3$ and $N_h=5$ have been used. The overlap factor will eventually
increase, for very large $\hat{R}$ according to (18).}
\label{fig9}
\end{figure}

\begin{figure}
 THE FIGURES CAN BE OBTAINED FROM THE AUTHORS
\end{figure}
\clearpage

\begin{center}
TABLE 1. Exciton energy $\hat{E}_0$ for different basis\\ set sizes.
Here $\hat{R}=2$, $m_e=m_h$, and $\hat{V}_e = \hat{V}_h = 5$.\\

\begin{tabular}{c|ccccc}                    \hline  \hline
$N_e \backslash N_h$ & 1 & 2 & 3 & 4 & 5\\  \hline
1 & 1.505 & 1.406 & 1.381 & 1.377 & 1.376\\
2 & 1.467 & 1.368 & 1.343 & 1.339 & 1.338\\
3 & 1.460 & 1.361 & 1.336 & 1.332 & 1.331\\
4 & 1.457 & 1.359 & 1.333 & 1.329 &      \\
5 & 1.456 & 1.357 & 1.331 &       &      \\ \hline \hline
\end{tabular}

\end{center}

\newpage

\begin{center}
{\large \sf Addendum}
\end{center}

\noindent
{\bf 1. Derivation of Eq.\ (15): Correlations for large mass ratio}\\

We seek the correlation function (14)
\begin{equation}
C=\frac{\langle \vec{r}_e\vec{r}_h \rangle}{\sqrt{\langle r_e^2\rangle\langle
r_h^2\rangle}}
\end{equation}
for the very special case of $V_e=\infty$, $V_h=0$ and  $m_e>>m_h$.

According to standard Born-Oppenheimer procedure one first solves the
ground-state Schr\"{o}dinger
 equation for the {\em light} particle (h), for fixed position
$\vec{r}_e$ of the heavy particle. This is a pure Coulombic problem yielding
\[ \psi_h(\vec{r}_h) = \frac{1}{\sqrt{\pi a_h^3}} \;e^{r/a_h}, \]
with $\vec{r} = \vec{r}_h-\vec{r}_e$. The eigenvalue for this problem is
 $-e^2/8\pi \varepsilon a_h$.

The second step in the Born-Oppenheimer procedure is to solve the
Schr\"{o}dinger equation for the heavy particle, with the eigenvalue
for the light-particle problem serving as an additional potential.
Since this eigenvalue is merely a constant, the confining potential
determines the wave function completely. The ground state is
 \[ \psi_e(r_e) =\frac{\sin(\pi r_e/R)}{\sqrt{2\pi R}\;\;r_e}.\]
The correlation function (1) is now easily evaluated using $\vec{r}$
instead of $\vec{r}_h$. With $\vec{r}_h = \vec{r}_e + \vec{r}$ we have
\begin{equation}
 C = \frac{\langle r_e^2+\vec{r} \;\vec{r}_e \rangle}{\sqrt{\langle r_e^2
\rangle \langle r_e^2 +2 \vec{r} \;\vec{r}_e + r^2 \rangle}} =
\sqrt{\frac{\langle r_e^2\rangle}{\langle r_e^2 + r^2\rangle }}
\label{CC}
\end{equation}
since $\langle \vec{r} \rangle = 0$ by symmetry.  The integrals
involved are simple:
\[ \langle r^2 \rangle = \frac{\int_0^{\infty} r^4 e^{-2r/a_h}\;
dr}{\int_0^{\infty} r^2 e^{-2r/a_h}\;dr} = 3 a_h^2, \]
and
\[ \langle r_e^2 \rangle =
 \frac{\int_0^R r_e^2 \sin^2(\pi r_e/R)\;
dr_e}{\int_0^R \sin^2(\pi r_e/R)\; dr_e} =
 R^2 \left(\frac{1}{3}-
\frac{1}{2\pi^2}\right). \]

Inserting these averages into (~\ref{CC}), and using $R = \hat{R} \;a_h$ we
obtain
\[ C = \frac{\hat{R}}{\sqrt{\hat{R}^2+9/(1-3/2\pi^2)}}, \]
which is equation (15) in the text.\\

\noindent
{\bf 2. Derivation of Eq.(16): The minimum width of the one-particle ground
state}\\

The unnormalized ground-state wave function $\psi(\vec{r})$ for
 one particle of mass $m$ in a dot potential
\begin{equation}
 V(\vec{r})  = \left\{ \begin{array}{cc}
                  0 & \mbox{$r < R$} \\
                  -V_e   & \mbox{$r \geq R$.}
                  \end{array}
                  \right.                              \label{pot}
\end{equation}
is given by
\begin{equation}
r\cdot \psi(r)  = \left\{ \begin{array}{cc}
                   \sin (k r/R) & \mbox{$r < R$} \\
                   \sin k e^{\beta (R-r)/R}     & \mbox{$r \geq R$.}
                  \end{array}
                  \right.                              \label{boelge}
\end{equation}
Here
\begin{equation}
k=R\sqrt{2mE/\hbar^2},
\label{k1}
\end{equation}
and
\begin{equation}
\beta =R\sqrt{2 m (V_e-E)/\hbar^2} =\sqrt{2mR^2V_e\hbar^{-2} - k^2}
\label{k2}
\end{equation}
with $E$ the ground-state energy.

The wave function (~\ref{boelge}) is continuous. Continuity of $\psi'(r)$
requires
\begin{equation}
\beta = -k \cot k.
\label{3}
\end{equation}

The width $<r^{2}>^{\frac{1}{2}}$ of the one-particle ground state is
thus given by
\begin{eqnarray}
  <r^{2}>&=&
   \frac{\int\limits_{0}^{R} r^{2} \sin^{2}(kr/R) dr
   +e^{2 \beta} \sin^{2}k\int\limits_{R}^{\infty } r^{2}
    e^{-2 \beta r/R } dr} {\int\limits_{0}^{R}  \sin^{2}(kr/R) dr
   +e^{2 \beta} \sin^{2}k\int\limits_{R}^{\infty } e^{-2 \beta r/R}
dr}\nonumber \\
& =& R^2 \frac{\frac{1}{6}-\frac{\sin(2k)}{4k}-\frac{\cos(2k)}{4k^2}
+\frac{\sin(2k)}{8k^3}+\frac{\sin^2k}{\beta}\left(\frac{1}{2}+\frac{1}{2\beta}
+\frac{1}{4\beta^2}\right)}{\frac{1}{2}-\frac{\sin(2k)}{4k}+\frac{\sin^2k}{2\beta}}
  \label{re2}
\end{eqnarray}

Using (~\ref{k2}) and (~\ref{3}) we have
\begin{equation}
R^2 = \frac{\hbar^2}{2mV_e}\cdot \frac{k^2}{\sin^2 k}.
\label{R}
\end{equation}
Thus we may express the width entirely in term of the variable $k$:
\begin{equation}
 <r^{2}>=\frac{\hbar^2}{2 m V_e}\cdot \frac{k^2}{\sin^2k}
 \frac{\frac{1}{6}-\frac{\sin (2k)}{4k}-\frac{\cos (2k)}{4k^2}
+\frac{\sin (2k)}{8k^3}-\frac{\sin^2k}{k\cot k}\left(\frac{1}{2}
-\frac{1}{2k\cot k}
+\frac{1}{4k^2\cot^2k}\right)}
{\frac{1}{2}-\frac{\sin (2k)}{4k}-\frac{\sin^2k}{2k\cot k}}
\label{k4}
\end{equation}
Thus
\begin{equation}
  r_{m}=\langle r^2\rangle^{\frac{1}{2}} =
  c \cdot\sqrt{\frac{\hbar^{2}}{2 m V_{e}}}.
\label{f}
\end{equation}
where c is a number.
 $c^2$ is the minimum value of the function of $k$ in (~\ref{f})
(after the multiplication sign).  Numerically
we find $c= 2.67$.

Finally, by introducing an effective Bohr radius as unit of length and an
effective
Rydberg as unit of energy, both in terms of the hole mass, equation (~\ref{f})
takes the dimensionless form
\begin{equation}
\hat{r}_{m}=c\;\sqrt{\frac{m_{h}}{m_{e}}}\sqrt\frac{1}{\hat{V}_{e}}.
\end{equation}
This is equation (16) in the manuscript.\\

\end{document}